\documentclass[aps,prc,showpacs,twocolumn,letter]{revtex4}

\usepackage{graphicx,epsfig}

\begin{document}

\title{Identified Particle Spectra and Anisotropic Flow in an Event-by-Event
Hybrid Approach in Pb+Pb collisions at $\sqrt{s_{\rm NN}}=2.76$ TeV}

\author{Hannah Petersen}
\affiliation{Department of Physics, Duke University, Durham, North Carolina
27708-0305, United States}


\begin{abstract}
The first results from heavy ion collisions at the Large Hadron Collider for
charged particle spectra and elliptic flow are compared to an event-by-event
hybrid approach with an ideal hydrodynamic expansion. This approach has been
shown to successfully describe bulk observables at RHIC. Without changing any
parameters of the calculation the same approach is
applied to Pb+Pb collisions at $\sqrt{s_{\rm NN}}=2.76$ TeV. This is an
important test if the established understanding of the dynamics of relativistic
heavy ion collisions is also applicable at even higher energies. Specifically,
we employ the hybrid approach with two different equations of state and the pure
hadronic transport approach to indicate sensitivities to finite viscosity. The
centrality dependence of the charged hadron multiplicity, $p_T$ spectra and
differential elliptic flow are shown to be in reasonable agreement with the
ALICE data.
Furthermore, we make predictions for the transverse mass spectra of identified
particles and triangular flow. The eccentricities and their fluctuations are
found to be surprisingly similar to the ones at lower energies and therefore
also the triangular flow results are very similar. Any deviations from these
predictions will indicate the need for new physics mechanisms responsible for
the dynamics of heavy ion collisions. 
\end{abstract}

\keywords{Relativistic Heavy-ion collisions, Monte Carlo simulations,
Hydrodynamic models}

\pacs{25.75.-q,24.10.Lx,24.10.Nz}

\maketitle

Recently, the first results from heavy ion collisions at $E_{\rm cm}=2.76A$ TeV
have been published by the ALICE collaboration
\cite{Aamodt:2010cz,Aamodt:2010jd,Aamodt:2010pa,Aamodt:2010pb}. To study
strongly interacting matter at high temperatures has been the goal of the
relativistic heavy ion program at the Relativistic Heavy Ion Collider (RHIC)
since more than a decade. The 10 times higher beam energies at the Large Hadron
Collider (LHC) allow for the investigation of the dynamical evolution of
nucleus-nucleus collisions that have been established in Au+Au collisions at
$E_{\rm cm}=200A$ GeV in a different kinematic range \cite{Abreu:2007kv}. It is
especially interesting, if the matter still behaves as a almost perfect liquid
or if the quark gluon plasma becomes more viscous by going to higher
temperatures \cite{Luzum:2010ag,Niemi:2011ix,Song:2011qa,Lacey:2010ej}. 

Hybrid approaches that are based on hydrodynamics for the hot and dense stage of
the evolution that is coupled to hadronic transport approaches to describe the
successive decoupling of the matter have been very successful in describing the
properties of the bulk matter that is created at RHIC
\cite{Bass:2000ib,Teaney:2001av,Hirano:2005xf,Nonaka:2006yn} and LHC
\cite{Hirano:2010je,Schenke:2011tv}. Within the last year, full event-by-event
hydrodynamic approaches
\cite{Andrade:2006yh,Tavares:2007mu,Petersen:2008dd,Holopainen:2010gz,
Werner:2010aa,Schenke:2010rr} have become more favorable because it has turned
out that the effect of initial state fluctuations on final flow observables
needs to be studied in a consistent way to draw quantitative conclusions for e.g
the value of the shear viscosity.   

It is important to apply well-established approaches for the dynamical evolution
of heavy ion reactions at RHIC to the higher energy collisions at LHC without
tuning parameters to investigate how good the energy dependence is described by
a specific model. Differences between the experimental data and the theoretical
calculations imply the need for new physics concepts to be applied. To
understand the bulk evolution at LHC energies is a pre-requisite for all further
detailed studies of e.g. jet quenching and energy loss or electromagnetic probes
\cite{Renk:2011gj,Holopainen:2011pd}. 

In this manuscript a event-by-event hybrid approach based on the
Ultra-relativistic Quantum Molecular Dynamics (UrQMD) transport approach with an
embedded (3+1) dimensional ideal hydrodynamic evolution is applied to lead-lead
collisions at LHC energies. First, a comparison to the available experimental
data on charged particle multiplicities, $p_{T}$ spectra and elliptic flow is
carried out with the exact same parameter set that has been applied at RHIC
energies. Then, transverse mass spectra and triangular flow for identified
particles are predicted. To compare the amount of initial state fluctuations,
the probability distributions of coordinate space eccentricities are compared to
their corresponding values at RHIC. 

Let us start with a short description of the hybrid approach that has been
developed for SPS energies \cite{Petersen:2008dd} and recently successfully
applied to gold-gold collisions at the highest RHIC energy
\cite{Petersen:2010di,Petersen:2010cw,Petersen:2010zt}. The early
non-equilibrium evolution is described by the UrQMD approach
\cite{Bass:1998ca,Bleicher:1999xi}, where the two lead nuclei are initialized
according to Wood-Saxon profiles followed by binary interactions of the
nucleons. The main contribution to the particle production at high energies is
achieved by string excitation and fragmentation where for the hard collisions
(momentum transfer $Q>1.5$ GeV) are treated by PYTHIA
\cite{Petersen:2008kb,NilssonAlmqvist:1986rx,Sjostrand:1993yb}. 

At the so called starting time of $t_{\rm start}=0.5$ fm the particle
distributions are transferred to energy, momentum and net baryon density
distributions by representing each particle with a three-dimensional Gaussian
distribution (width $\sigma=1$ fm) that is Lorentz-contracted along the beam
direction. During the following ideal relativistic one fluid evolution
\cite{Rischke:1995ir,Rischke:1995mt} two different equations of state are
employed, one representing a hadron gas (HG-EoS) and one including a cross-over
deconfinement phase transition based on a chiral approach
(DE-EoS)\cite{Steinheimer:2009nn,Steinheimer:2009hd}. The transition back to the
hadronic transport approach happens on a constant proper time hypersurface,
where the Cooper-Frye equation is applied on transverse slices of thickness
$\Delta z=0.1-0.2$ fm that have cooled down below an energy density of $5
\epsilon_0\approx 730$ MeV/fm$^3$ \cite{Petersen:2009mz}. This approach provides
the full final phase space distributions of the produced particles for each
event and can be compared to the pure transport approach by turning off the
hydrodynamic evolution which allows for a qualitative study of viscous effects.

\begin{figure}[ht]
\resizebox{0.5\textwidth}{!}{ \centering
\includegraphics{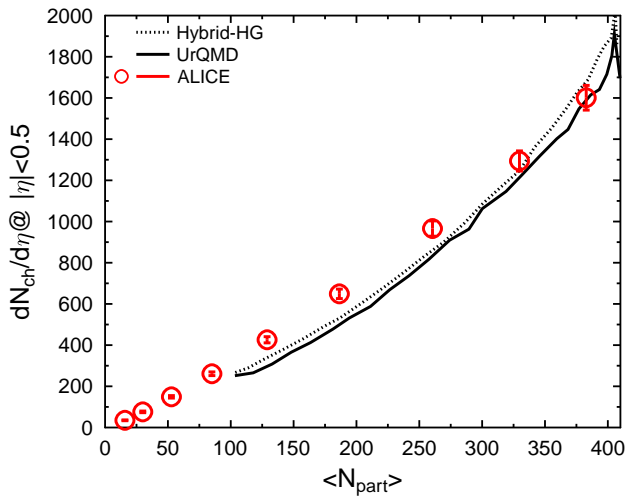}
}
\caption{(Color online) Charged particle multiplicity at midrapidity
($|\eta|<0.5$) as a function of the number of participants in Pb+Pb collisions
at $\sqrt{s_{\rm NN}}=2.76$ TeV calculated in the UrQMD transport and the hybrid
approach compared to the experimental
data \cite{Aamodt:2010cz}.} 
\label{fig_mulimpch}
\end{figure}

The first observable to look at is the charged particle multiplicity at
midrapidity. In Fig. \ref{fig_mulimpch} the calculation of the centrality
dependent multiplicity scaled by the number of participants (estimated in a
Glauber approach) is shown. The hadronic transport approach UrQMD provides a
reasonable description of the multiplicity. For central collisions the
predictions published in \cite{Mitrovski:2008hb} are right on top of the ALICE
data while with decreasing centrality the number of charged particles is a
little
lower than in the data. This fair agreement with the data hints to the fact that
the main particle production can be described by the initial binary
nucleon-nucleon interactions treated by PYTHIA. The hydrodynamic evolution does
not affect the particle production. Since ideal hydrodynamics implies an
isentropic expansion this means that the charged particle multiplicity is
determined 
in the initial state and by the final resonance decays.

\begin{figure}[ht]
\resizebox{0.5\textwidth}{!}{ \centering
\includegraphics{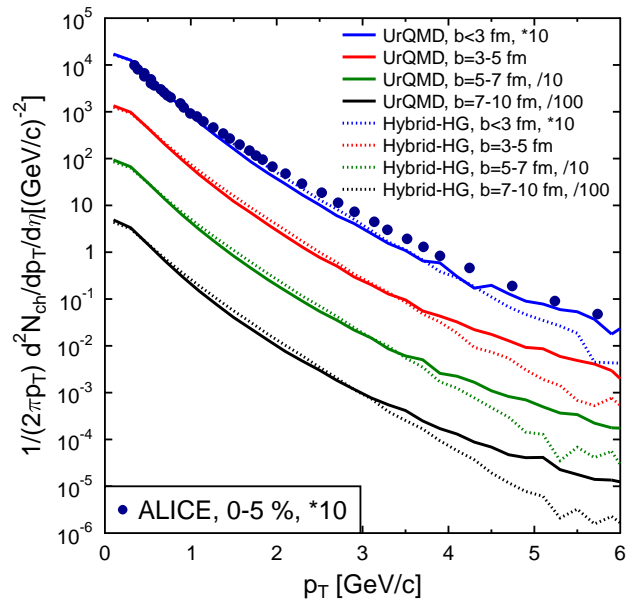}
}
\caption{(Color online) Transverse momentum spectra of charged particles for
four different centralities calculated in the UrQMD transport and the hybrid
approach compared to the available experimental data \cite{Aamodt:2010jd}.} 
\label{fig_dndptimpch}
\end{figure}

For the following calculations of spectra and collective flow four different
centrality classes have been chosen that match the ones applied by the ALICE
collaboration as they are listed in the  following table:

\vspace{0.5cm}
\begin{tabular}{|c|c|}
\hline
Centrality class&Impact parameter range\\
\hline
0-5\%&$b<3$ fm\\
5-10\%&$b=3-5$ fm\\
10-20\%&$b=5-7$ fm\\
20-40\%&$b=7-10$ fm\\
\hline
\end{tabular}
\vspace{0.5cm}

The transverse momentum spectrum for charged particles in the mentioned
centrality classes are compared to experimental data in the most central bin
(see Fig. \ref{fig_dndptimpch}). The main difference between the hybrid and the
transport calculation is in the slopes of the spectra. As expected the
hydrodynamic evolution leads to a purely exponential $p_T$ dependence which
describes the data until $p_T<3$ GeV very well. At higher transverse momenta the
power law tail from hard processes becomes important for a good agreement with
the measured values. In the range from 4 to 6 GeV the non-equilibrium
description exemplified by the UrQMD calculation provides a better description
of the experimental data.

\begin{figure}[ht]
\resizebox{0.5\textwidth}{!}{ \centering
\includegraphics{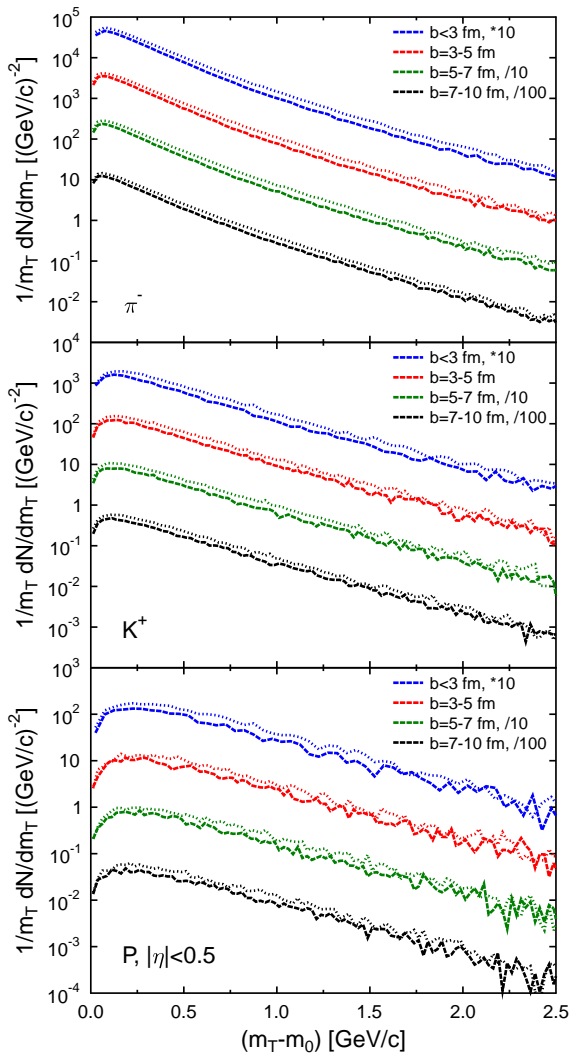}
}
\caption{(Color online) Transverse mass spectra of negative pions (top),
positive kaons (middle) and protons (bottom) for
four different centralities calculated in the hybrid
approach with two different equations of state.} 
\label{fig_dndmtimpall}
\end{figure}

In Fig. \ref{fig_dndmtimpall} predictions for the transverse mass spectra at
midrapidity of pions, kaons and protons are presented. The pion spectra are very
similar to the charged particle spectra since they represent the major fraction
of the newly produced particles in the collision. Kaons are strange mesons and
protons are chosen because they have a higher mass and are baryonic degrees of
freedom. The general features of the transverse mass spectra are similar to the
ones observed at RHIC and imply a collective radial velocity that drives all the
particle species. The two different equations of state lead to very similar
results with the deconfinement transition having a little steeper slope due to
the more rapid expansion due to the higher pressure in the quark gluon plasma
phase.  

\begin{figure}[ht]
\resizebox{0.5\textwidth}{!}{ \centering
\includegraphics{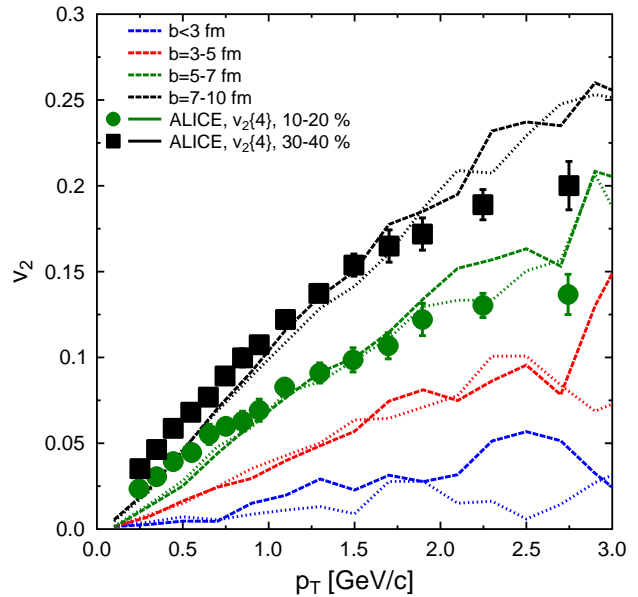}
}
\caption{(Color online) Elliptic flow of charged particles as a function of
transverse momentum for
four different centralities calculated in the hybrid
approach with two different equations of state compared to the experimental
data\cite{Aamodt:2010pa}.} 
\label{fig_v2ptimpch}
\end{figure}

After proving a rather successful agreement with basic quantities like the
multiplicity and transverse momentum spectrum the next step is to look at 
anisotropic flow observables. The elliptic flow has been calculated with respect
to the reaction plane by averaging over all charged particles in all events to
be compared to the ALICE measurement that relies on the four-particle cumulant
method in two centrality bins. Fig. \ref{fig_v2ptimpch} shows a good agreement
between the hybrid calculations and the data, especially between $p_T$=0.8-2.5
GeV. In the very low transverse momentum region the hybrid approach
underpredicts the data which has been observed in other calculations as well
\cite{Hirano:2010je}. At
higher $p_T$ again the influence of hard processes needs to be taken into
account.

\begin{figure}[ht]
\resizebox{0.5\textwidth}{!}{\centering
\includegraphics{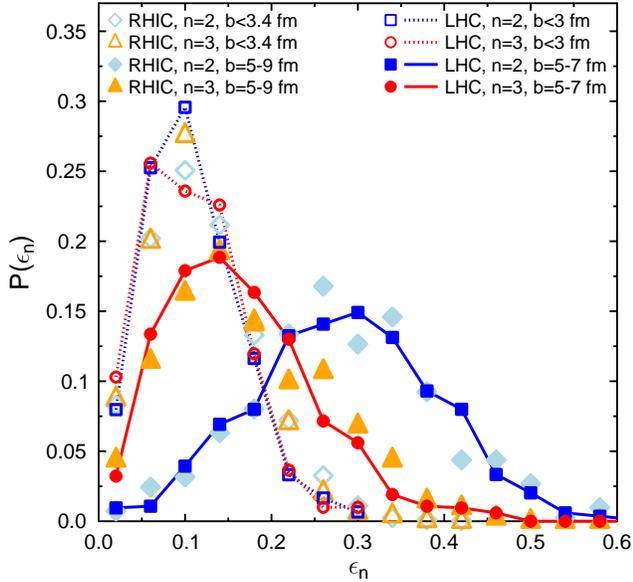}
}
\caption{(Color online) Probability distributions of the eccentricity
coefficients ($\epsilon_2/\epsilon_3$) in the UrQMD initial state.The diamonds
and triangles indicate the corresponding distributions for Au+Au collisions at
RHIC at
$\sqrt{s_{\rm NN}}=200$ GeV.} 
\label{fig_probeps}
\end{figure}

To quantify the shape of the initial conditions employed for the hydrodynamic
calculation and its event-by-event fluctuations Fig. \ref{fig_probeps} shows the
probability distribution of the coordinate space asymmetry characterized by the
eccentricity and the triangularity as defined in \cite{Petersen:2010cw}. The
initial $\epsilon_n$ coefficients have been calculated in each event and the
normalized probability distribution is plotted for two different centrality
bins. 

For central collisions the mean value and the shape of the distributions are
very similar for the participant eccentricity and the triangularity since both
of them are mainly generated by fluctuations. For more peripheral collisions the
eccentricity is influenced by a large geometry component due to the ellipsoidal
shape of the initial state in the transverse plane. Therefore, the mean
eccentricity is larger and the fluctuations increase leading to a wider
distribution,
whereas the triangularity stays smaller and the distribution has a smaller
width. 

Since the triangularity has been introduced because of its sensitivity to
initial state fluctuations the higher multiplicity at LHC energies triggers the
expectations that the fluctuations become smaller compared to RHIC energies. In
Fig. \ref{fig_probeps} the triangles and diamonds depict the eccentricity and
triangularity calculation from UrQMD initial conditions for Au+Au collisions at
$E_{\rm cm}=200A$ GeV. Surprisingly, the $\epsilon_n$ distributions match almost
exactly the ones at LHC energies for the two similar centrality classes. 

\begin{figure}[ht]
\resizebox{0.5\textwidth}{!}{ \centering
\includegraphics{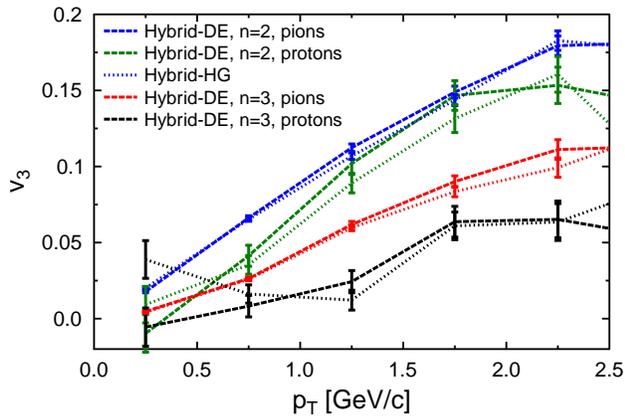}
}
\caption{(Color online) Triangular flow of pions and protons as a function of
transverse momentum compared to the corresponding elliptic flow result
calculated in the hybrid approach with two different equations of state for
minimum bias Pb+Pb collisions at $\sqrt{s_{\rm NN}}=2.76$ TeV.}
\label{fig_v3ptid}
\end{figure}

To calculate the transverse momentum dependence of the anisotropic flow
coefficients for identified particles (see Fig. \ref{fig_v3ptid}) the event
plane method including removal of auto-correlations and resolution correction
has been applied as described in more detail in \cite{Petersen:2010cw}. In this
way, the
predictions can be directly compared to a future experimental result for minimum
bias Pb+Pb collisions at $E_{\rm cm}=2.76A$ TeV. Both elliptic and triangular
flow show the expected mass splitting between pions and protons. Furthermore,
the elliptic flow has roughly double the value of the triangular flow result as
it has previously been calculated for RHIC energies. The anisotropic flow are
not very sensitive to the choice of the equation of state which points to the
fact that the interplay of freeze-out transition and flow development due to
pressure gradients results in very similar values independent of the
existence of a quark gluon plasma phase. 

\begin{figure}[ht]
\resizebox{0.5\textwidth}{!}{ \centering
\includegraphics{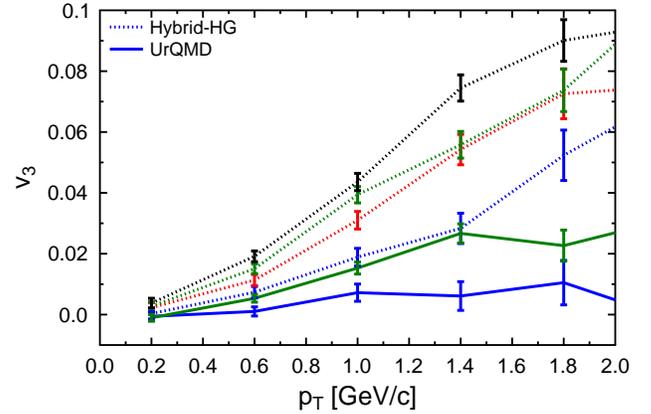}
}
\caption{(Color online) Triangular flow of charged particles as a function of
transverse momentum calculated in the hybrid approach for four different
centrality classes of Pb+Pb collisions at
$\sqrt{s_{\rm NN}}=2.76$ TeV compared in two centrality bins to the
corresponding results of the UrQMD
transport approach. }
\label{fig_v3ptch}
\end{figure}

Finally in Fig. \ref{fig_v3ptch} predictions for the transverse momentum
dependence of triangular flow for charged particles in the four different
centrality classes are made. There is a weak centrality dependence, but overall
the different curves from the hybrid calculation look very similar
qualitatively. Since the calculation with the deconfinement phase transition
leads to very similar results, only the hadron gas result is shown in this
figure to reduce the number of lines displayed. To investigate the
effect of a finite viscosity during the expansion the calculation is compared to
the pure UrQMD result. Within the very viscous hadronic transport approach the
pressure gradients are too small to transfer the initial coordinate space
asymmetry to a final state momentum space asymmetry. For impact parameters up to
$b=5$ fm the result is consistent with zero and even for very peripheral event
the result is negligible especially in the low transverse momentum region where
most of the particles are. This leads to the conclusion that a finite triangular
flow measurement is a strong indication for almost ideal hydrodynamic behavior
during the hot and dense evolution of the heavy ion reaction. 

To summarize we have presented a comparison of bulk observables measured in
Pb+Pb collisions at $E_{\rm cm}=2.76A$ TeV to a state-of-the-art event-by-event
hybrid description. Employing the same parameters as have been used for Au+Au
collisions at RHIC a reasonably good agreement for the multiplicity, transverse
momentum spectra and elliptic flow is achieved. One can conclude that
the basic assumptions about the major ingredients that are needed to describe
the dynamic evolution of heavy ion reactions do not have to be revised at LHC
energies. 

With a non-equilibrium initial state, an ideal hydrodynamic evolution and a
hadronic afterburner predictions for identified particle transverse mass
spectra and elliptic as well as triangular flow are made. The initial state
fluctuations
quantified by the probability distribution of the initial eccentricity and
triangularity are very similar at LHC and RHIC energies within the UrQMD
approach. A large viscosity during the evolution results in almost negligible
higher harmonic coefficients, so the triangular flow measurement at LHC will
provide a robust confirmation of the almost ideal hydrodynamic expansion. 

\section*{Acknowledgements}
\label{ack} We are grateful to the Open Science Grid for the computing
resources. The author thanks Dirk Rischke for
providing the 1 fluid hydrodynamics code. H.P. acknowledges a Feodor Lynen
fellowship of the Alexander von Humboldt
foundation. This work was supported in part by U.S. department of Energy grant
DE-FG02-05ER41367. H.P. thanks Jan Steinheimer
for help with the extension of the equation of state, Guangyou Qin for providing
the Glauber calculation of the numebr of particpants and Steffen
A. Bass and Berndt M\"uller for
fruitful discussions. 


\end{document}